\documentclass{article}
\usepackage{amsmath}

\input{tcilatex}
\begin{document}

\title{The quantum and Klein-Gordon oscillators in a non-commutative complex
space and the thermodynamic functions}
\author{S. Zaim \\
D\'{e}partment des Sciences de la Mati\`{e}re, Facult\'{e} des Sciences,\\
Universit\'{e} Hadj Lakhdar - Batna, Algeria}
\date{}
\maketitle

\begin{abstract}
In this work we study the quantum and Klein-Gordon oscillators in
non-commutative complex space. We show that the quantum and Klein-Gordon
oscillators in non-commutative complex space are obeys an particle similar
to the an electron with spin in a commutative space in an external uniform
magnetic field. Therefore the wave function $\psi \left( z,\bar{z}\right) $
takes values in $C^{4}$, spin up, spin down, particle, antiparticle, a
result which is obtained by the Dirac theory. The energy levels could be
obtained by exact solution. We also derived the thermodynamic functions
associated to the partition function, and we show that the non-commutativity
effects are manifested in energy at the high temperature limit.
\end{abstract}

\section{Introduction}

In recent years many arguments have been suggested to motivate a deviation
from the flat-space concept at very short distances $\left[ 1,2\right] $ so
that we have a new concept of quantum spaces $\left[ 3,4,5,6,7\right] $.
Quantum spaces depend on parameters such that for a particular value of
these parameters they become the usual flat space. We consider the simplest
version of quantum spaces as the natural extension of the usual quantum
mechanical commutation relations between position and momentum, by imposing
further commutation relations between position coordinates themselves. The
non-commutative space can be obtained by the coordinate operators where we
replace the ordinary product by the star product: 
\begin{equation}
\left[ x^{\mu },x^{\nu }\right] _{\star }=i\theta ^{\mu \nu },
\end{equation}
or using the new non-commutative coordinate operators $\hat{x}$ satisfying
the commutation relations: 
\begin{equation}
\left[ \hat{x}^{\mu },\hat{x}^{\nu }\right] =i\theta ^{\mu \nu },
\end{equation}
where 
\begin{equation}
\hat{x}^{\mu }=x^{\mu }-\frac{\theta ^{\mu \nu }}{2}p_{\nu }.
\end{equation}

For the non-commutative canonical-type space, the parameter $\theta ^{\mu
\nu }$ is an antisymmetric real matrix of dimension length-square. The
star-product $\left( \star \right) $ defined between two function is given
by: 
\begin{equation}
\psi \left( x\right) \star \varphi \left( x\right) =\psi \left( x\right)
\exp \left\{ -i\theta ^{\mu \nu }\overleftarrow{\partial }_{\mu } 
\overrightarrow{\partial }_{\nu }\right\} \varphi \left( y\right) \mid
_{x=y}.
\end{equation}
It is easy to verify the relation (1) by simply replacing $\psi \left(
x\right) =x^{\mu },\varphi \left( x\right) =x^{\nu }$ in $(4)$. This idea is
similar to the physical meaning of the Plank constant in the relation $%
[x,p]=i\hbar $, which as is known is the smallest phase-space in quantum
mechanics. There is a lot of interest in recent years in the study of
non-commutative canonical-type quantum mechanics, quantum field theory and
string theory $\left[ 8,9,10\right]$. On other hand the solutions of
classical dynamical problems of physical systems obtained in terms of
complex space variables are well-known. There are also interests in the
complex quantum mechanic systems (in two dimentions) $\left[ 11,12\right]$,
in which we consider a quantum harmonic oscillator in non-commutative
complex quantum space (so coordinate and momentum operators of this space
are written $\hat{z}=\hat{x}+i\hat{y}$ and $p_{\hat{z}}=\left(
p_{x}-ip_{y}\right) /2$).

In this paper we first present a harmonic oscillator in the non-commutative
complex quantum space. The system is described by the wave function $\psi
\left( z,\bar{z}\right) $ which takes values in $C^{4}$, spin up, spin down,
particle, antiparticle, which is obtained by the Dirac formalism. Secondly
we shall also study relativistic quantum mechanics particularly we show that
the Klein-Gordon oscillator in the non-commutative complex quantum space is
similar to the equation of motion for a fermion with spin $1/2$ $\left[ 13%
\right]$.

This paper is organized as follows. In section $2$ we present the quantum
oscillator in a non-commutative complex space. We shall see the effect of
non-commutativity of space on the thermodynamics function associated to the
harmonic oscillator with a non-commutative complex space, and we show how
our findings differ from the results obtained in ref $\left[ 14,15\right]$.
In section $3$ we derive the deformed Klein-Gordon equation for the harmonic
oscillator with non-commutative complex space. We solve this equation and
obtain the non-commutative modification of the energy levels. Finally, in
section $5$, we draw our conclusions.

\section{ Quantum oscillator in non-commutative complex space}

In two dimensional space, the complex coordinate system $\left( z,\bar{z}
\right)$ and momentum $\left( p_{z},p_{\bar{z}}\right)$ is defined by: 
\begin{equation}
z=x+iy,\quad \bar{z}=x-iy, \quad \text{and} \quad p_{z}=\frac{1}{2}\left(
p_{x}-ip_{y}\right),\quad p_{\bar{z}}=-\bar{p}_{z}=\frac{1}{2} \left(
p_{x}+ip_{y}\right).
\end{equation}

We are interested in introducing the non-commutative complex operators of
coordinates and momentum in a two-dimensional space: 
\begin{eqnarray}
\hat{z} =\hat{x}+i\hat{y}=z+i\theta p_{\bar{z}},&\qquad& \widehat{\bar{z } }
=\hat{x}-i\hat{y}=\bar{z}-i\theta p_{z}, \\
\hat{p}_{z} =p_{z},&\qquad& \hat{p}_{\bar{z}}=p_{\bar{z}}.
\end{eqnarray}
The non-commutative algebra $(1)$ can be rewritten as: 
\begin{equation}
\left[\hat{z},\hat{\bar{z}}\right] =2\theta,\, \left[ \hat{z},p_{\hat{z}} %
\right]=\left[ \hat{\bar{z}},p_{\hat{z}}\right] =0,\,\left[ \hat{z},p_{\hat{z%
}}\right] =\left[ \hat{\bar{z}}, p_{\hat{z}} \right] = 2\hbar,\, \left[ p_{%
\hat{z}},p_{\hat{z}}\right] =0.
\end{equation}

Now we will discuss the oscillator systems on non-commutative complex
quantum space. In this formulation we consider the isotropic oscillator on a
two dimensional space: 
\begin{eqnarray}
H &=&\frac{1}{2m}\left( p_{x}^{2}+p_{y}^{2}\right) +\frac{m\omega ^{2}}{2}%
\left( x^{2}+y^{2}\right)  \notag \\
&=&\frac{1}{2m}\left( p_{x}-ip_{y}\right) \left( p_{x}+ip_{y}\right) +\frac{%
m\omega ^{2}}{2}\left( x+iy\right) \left( x-iy\right) \,.
\end{eqnarray}%
Using the eqs. (5), eq. (9) takes the form: 
\begin{equation}
H=H_{z\bar{z}}=\frac{2}{m}p_{z}p_{\bar{z}}+\frac{m\omega ^{2}}{2}z\bar{z}=H_{%
\bar{z}z}.
\end{equation}%
This Hamiltonian of an oscillator on a complex space is Hermitian. The
solution of the corresponding eigenvalue is real. We can introduce the new
annihilation operators $a$ and $b$ defined as: 
\begin{eqnarray}
a &=&\frac{2ip_{\bar{z}}+m\omega z}{2\sqrt{m\omega }}, \\
b &=&\frac{2ip_{z}+m\omega \bar{z}}{2\sqrt{m\omega }},
\end{eqnarray}%
and the corresponding creation operators $a^{+}$ and $b^{+}$ satisfying the
usual commutation relations: 
\begin{equation}
\left[ a,a^{+}\right] =\left[ b,b^{+}\right] =1.
\end{equation}%
All other commutations are zero. We can express $z,\bar{z}$ and $p_{z},p_{%
\bar{z}}$ in terms of these operators as: 
\begin{eqnarray}
p_{z} &=&\frac{i}{2}\sqrt{m\omega }\left( -a+b^{+}\right) , \\
\bar{p}_{z} &=&\frac{i}{2}\sqrt{m\omega }\left( a^{+}-b\right) , \\
z &=&\frac{1}{\sqrt{m\omega }}\left( a+b^{+}\right) , \\
\bar{z} &=&\frac{1}{\sqrt{m\omega }}\left( b+a^{+}\right) .
\end{eqnarray}%
After replacing $(14)$ into $(17)$ the Hamiltonian is written as: 
\begin{equation}
H_{z\bar{z}}=\omega \left( a^{+}a+b^{+}b+1\right) .
\end{equation}

The operators $a^{+}a=N^{+}$ and $b^{+}b=N_{+}$ satisfy the eigenvalue
equations $N_{+}^{+}\left\vert n^{+},n_{+}\right\rangle =n_{+}^{+}\left\vert
n^{+},n_{+}\right\rangle $ with $n_{+}^{+}=0,1,2,...$. The eigenvalues for
the Hamiltonian $(18)$ are given by: 
\begin{equation}
E_{n^{+}n_{+}}=\omega \left( n^{+}+n_{+}+1\right).
\end{equation}
To calculation the wave functions $\psi _{n^{+}n_{+}}\left( z,\bar{z}\right) 
$, we simply apply the operators $a^{+}$ and $b^{+}$ on the ground state $%
\psi _{00}\left( z,\bar{z}\right) $ which is give by: 
\begin{equation}
\psi _{00}\left( z,\bar{z}\right) =\sqrt{\frac{m\omega }{\pi }}\exp \left( -%
\frac{m\omega }{2}z\bar{z}\right).
\end{equation}
We see from eq. $(19)$ that the eigenvalues of $H_{z\bar{z}}$ take the form: 
\begin{equation}
E_{n}=\omega \left( n+1\right),
\end{equation}
where 
\begin{equation}
n=n^{+}+n_{+}.
\end{equation}
For each value of the energy (labeled by $n$), the wave functions of the
system is described by two wave functions $\left[ 16,12\right]$: 
\begin{equation}
\psi_{n0}\left( z,\bar{z}\right) = Cz^{n}\exp \left( -\frac{m\omega }{2}z 
\bar{z}\right) ,\qquad \psi _{0n}\left( z,\bar{z}\right) = C\bar{z}^{n}\exp
\left( - \frac{m\omega }{2}z\bar{z}\right)
\end{equation}
where $C=\sqrt{\frac{\left( m\omega \right) ^{n+1}}{\pi n!}}$

In the non-commutative complex space we notice that $\hat{z}\hat{\bar{z}}
\neq \hat{\bar{z}}\hat{z}$. Then the Hamiltonian of equation (10) can be
written as: 
\begin{equation}
\hat{H}=\left( 
\begin{array}{cc}
\hat{H}_{\hat{z}\hat{\bar{z}}} & 0 \\ 
0 & \hat{H}_{\hat{\bar{z}}\hat{z}}%
\end{array}
\right),
\end{equation}
where 
\begin{equation}
\hat{H}_{\hat{z}\hat{\bar{z}}}=\frac{2}{m}p_{\hat{z}}p_{\hat{\bar{z}} }+%
\frac{m\omega ^{2}}{2}\hat{z}\hat{\bar{z}},
\end{equation}
and 
\begin{equation}
\hat{H}_{\hat{\bar{z}}\hat{z}}=\frac{2}{m}p_{\hat{z}}p_{ \hat{\bar{z}}}+%
\frac{m\omega ^{2}}{2}\hat{\bar{z}}\hat{z}.
\end{equation}

The Hamiltonian of equation $(25)$ takes the form: 
\begin{equation}
\hat{H}_{\hat{z}\hat{\bar{z}}}=\frac{2}{\tilde{m}}p_{z}p_{\bar{z}}+\frac{1}{2%
}\tilde{m}\tilde{\omega}^{2}z\bar{z}-\frac{m\omega ^{2}}{2}\theta \left(
L_{z}-1\right),
\end{equation}
where $\frac{1}{\tilde{m}}=\frac{1}{m}+\frac{m\omega ^{2}}{2}\theta ^{2}$, $%
\tilde{\omega}=\omega \sqrt{1+\frac{m^{2}\omega ^{2}}{2}\theta ^{2}}$ and $%
L_{z}=i\left( zp_{z}-\bar{z}p_{\bar{z}}\right) $. The last term can be
written as: 
\begin{equation}
-\frac{m\omega ^{2}}{2}\theta \left( L_{z}-1\right) =-\frac{m\omega ^{2}}{2}
\theta \left( L_{z}+2s_{z}\right),\quad \text{where}\quad s_{z}=-1/2,
\end{equation}
which is the same as the total magnetic moment of particle with spin $1/2$
where the non-commutativity parameter plays the role of a magnetic field.

Then the Hamiltonian of eq. $(27)$ takes the form: 
\begin{equation}
\hat{H}_{\hat{z}\hat{\bar{z}}}=\frac{2}{\tilde{m}}p_{z}p_{\bar{z}}+\frac{ 1}{%
2}\tilde{m}\tilde{\omega}^{2}z\bar{z}-\frac{m\omega ^{2}}{2} \theta \left(
L_{z}+2s_{z}\right)=\hat{H}_{\downarrow },\quad \text{where}\quad s_{z}=-1/2.
\end{equation}
This Hamiltonian is Hermitian and withe represents the oscillation of a
particle with spin $1/2$ in a uniform external magnetic field where the
direction of spin is opposite to that of the magnetic field.

Furthermore the Hamiltonian of equation $(26)$ takes the form: 
\begin{equation}
\hat{H}_{\hat{\bar{z}}\hat{z}}=\frac{2}{\tilde{m}}p_{z}p_{\bar{z}}+\frac{ 1}{%
2}\tilde{m}\tilde{\omega}^{2}z\bar{z}-\frac{m\omega ^{2}}{2} \theta \left(
L_{z}+2s_{z}\right) =\hat{H}_{\uparrow },\quad \text{where} \quad s_{z}=1/2.
\end{equation}
This Hamiltonian is Hermitian and represents a particle with spin $1/2$ in a
uniform external magnetic field where the projection of spin is along the
direction of the magnetic field.

The Hamiltonian in eqs. $(29)$ and $(30)$ is Hermitian and can written in
terms of creation/annihilation operators as: 
\begin{equation}
\hat{H}=\left( 
\begin{array}{cc}
\begin{array}{c}
\tilde{\omega}\left( a^{+}a+b^{+}b+1\right) + \\ 
\frac{m\omega ^{2}}{2}\theta \left( a^{+}a-b^{+}b+1\right)%
\end{array}
& 0 \\ 
0 & 
\begin{array}{c}
\tilde{\omega}\left( a^{+}a+b^{+}b+1\right) + \\ 
\frac{m\omega ^{2}}{2}\theta \left( a^{+}a-b^{+}b-1\right)%
\end{array}%
\end{array}%
\right).
\end{equation}
Here the annihilation operators $a$ and $b$ are 
\begin{eqnarray}
a &=&\frac{2p_{\bar{z}}-i\tilde{m}\tilde{\omega}z}{2\sqrt{\tilde{m}\tilde{%
\omega}}}, \\
b &=&\frac{2p_{z}-i\tilde{m}\tilde{\omega}\bar{z}}{2\sqrt{\tilde{m}\tilde{%
\omega}}},
\end{eqnarray}
and we have replaced the angular momentum $L_{z}$ in terms of $a,a^{+}$ and $%
b,b^{+}$ through: 
\begin{equation}
L_{z}=-\left( a^{+}a-b^{+}b\right).
\end{equation}

The eigenvalues for the Hamiltonian $(31)$ are: 
\begin{equation}
E_{\pm }=\tilde{\omega}\left( n^{+}+n_{+}+1\right) +\frac{m\omega ^{2}}{2}%
\theta \left( n^{+}-n_{+}\pm 1\right).
\end{equation}
We see from eq. (35) that the eigenvalues of $\hat{H}$ take the form: 
\begin{equation}
E_{n}=\tilde{\omega}\left( n+1\right) +\frac{m\omega ^{2}}{2}\theta \left(
m_{l}\pm 1\right),
\end{equation}
where $n=n^{+}+n_{+}$ and $m_{l}=n^{+}-n_{+}$. For a given value of the
energy labeled by $n$, the values of $m_{l}$ are bound by $\left( n\right) $
from above and by $\left( -n\right) $ from below. The wave function of the
system is described by four wave functions: 
\begin{eqnarray}
\psi _{n0}^{+}\left( z,\bar{z}\right) =Cz^{n}\exp \left( -\frac{m\omega }{2 }%
z\bar{z}\right) \left\vert \uparrow \right\rangle , \psi _{n0}^{-}\left( z, 
\bar{z}\right) =Cz^{n}\exp \left( -\frac{m\omega }{2}z\bar{z}\right)
\left\vert \downarrow \right\rangle && \\
\psi _{0n}^{-}\left( z,\bar{z}\right) =C\bar{z}^{n}\exp \left( -\frac{
m\omega }{2}z\bar{z}\right) \left\vert \downarrow \right\rangle , \psi
_{0n}^{+}\left( z,\bar{z}\right) =C\bar{z}^{n}\exp \left( -\frac{m\omega }{2}
z\bar{z}\right) \left\vert \uparrow \right\rangle&&
\end{eqnarray}
which correspond to the limiting values of $L_{z}$ being maximal $\left(
n\right) $ and minimal $\left( -n\right)$. Thus a single oscillator state
may be split into two pairs each having the same energy leading to twofold
degeneracy of the energy levels. This oscillator is described by two double
component spinor: 
\begin{equation}
\psi _{n0} =\left( 
\begin{array}{c}
\psi_{n0}^+ \\ 
\psi_{n0}^-%
\end{array}%
\right),\qquad \text{and}\qquad\psi _{0n}=\left(%
\begin{array}{c}
\psi _{0n}^{+} \\ 
\psi _{0n}^{-}%
\end{array}%
\right) ,
\end{equation}
where the sign $\left( \pm \right) $ signifies spin up or down. The
oscillator is positioned in the four equivalent points $\left( z_{\uparrow
}, \bar{z}_{\uparrow },z_{\downarrow },\bar{z}_{\downarrow }\right)
\Leftrightarrow \left( z,\bar{z},-z,-\bar{z}\right) $. Therefore the wave
function $\psi \left( z,\bar{z}\right) $ takes values in $C^{4}$, spin up,
spin dwon, particle, antiparticle. This result is obtained by the Dirac
theory.

For the special case $n^{+}=n_{+}=n$ the eigenvalues are: 
\begin{equation}
E_{\pm }=\tilde{\omega}\left( 2n+1\right) \pm \frac{m\omega ^{2}}{2}\theta\,
.
\end{equation}
They correspond to the energy values for the charged oscillator with spin $%
\frac{ 1}{2}$ in a magnetic field, where the non-commutativity plays the
role of magnetic field which creates the total magnetic moment of a particle
with spin $1/2$, which in turn shifts the spectrum of energy resulting in
the degeneracy being removed. Such effects are similar to the Zeeman
splitting in a commutative space.

We note that the Hamiltonian of harmonic oscillator in non-commutative
ordinary coordinates reads: 
\begin{equation}
\hat{H}=\frac{1}{2}\left( \hat{H}_{\downarrow }+\hat{H}_{\uparrow }\right) =%
\frac{1}{2m}\left( p_{x}^{2}+p_{y}^{2}\right) +\frac{m\omega ^{2}}{2}\left( 
\hat{x}^{2}+\hat{y}^{2}\right).
\end{equation}
The eigenvalues of this Hamiltonian can be written as: 
\begin{equation}
E=\tilde{\omega}\left( n^{+}+n_{+}+1\right) +\frac{m\omega ^{2}}{2}\theta
\left( n^{+}-n_{+}\right)
\end{equation}
For the special case $n^{+}=n_{+}=n$, the energy $E=\tilde{\omega}\left(
2n+1\right) $ is not shifted, contrary to non-commutative complex space
where the energy levels are shifted. Thus the system without spin in
non-commutative coordinate space has an added advantage that the spin effect
is automatically manifested.

The thermodynamic functions associated with the non-commutative complex
oscillator are also of interest. First we calculate the partition function $%
Z\left( \beta ,\theta \right)$ $\left[ 14,15\right] $: 
\begin{equation}
Z\left( \beta ,\theta \right) =trG\left( x,\acute{x}\right),
\end{equation}
where $G\left( x,\acute{x}\right) $ is the Green function given by: 
\begin{equation}
G\left( x,\acute{x};T\right) =\left\langle x\right\vert \exp \left\{
-i\left( \hat{H}\right) \left( t-\acute{t}\right) \right\} \left\vert \acute{
x}\right\rangle.
\end{equation}
The Hamiltonian $\hat{H}$ is diagonal in the basis $\left\vert
n^{+}n_{+}\right\rangle $ and the energy eigenvalue given in the
relationship $(35)$ can be expressed by: 
\begin{equation}
E_{\pm }=\hbar \tilde{\omega}\left( \left( 1+\frac{m\omega }{2}\theta
\right) n^{+}+\left( 1-\frac{m\omega }{2}\theta \right) n_{+}+\left( 1\pm 
\frac{m\omega }{2}\theta \right) \right).
\end{equation}
Then the Green function in eq. $(44)$ is readily found to be: 
\begin{multline}
G\left( x,\acute{x};T\right) =\sum_{n^{ {{}^+} }n_{+}}\psi _{n^{ {{}^+}
}n_{+}}^{+}\left( x\right) \psi _{\acute{n}^{ {{}^+} }\acute{n}_{+}}\left( 
\acute{x}\right) \times \\
\times \exp \left\{ -i\left( \hbar \tilde{\omega}\left( \left( 1+\frac{%
m\omega }{2 }\theta \right) n^{+}+\left( 1-\frac{m\omega }{2}\theta \right)
n_{+}+\left( 1+\frac{m\omega }{2}\theta \right) \right) \right) T\right\},
\end{multline}
where $T=t-\acute{t}$ and $\psi _{n^{ {{}^+}}n_{+}}\left( x\right) $ are the
eigenfunctions of the harmonic oscillator in in commutative space. Using the
Euclidean rotation $i\left( t-\acute{t} \right) \rightarrow \beta $, one can
see that the partition function in equation $(43)$ can be written by as: 
\begin{eqnarray}
Z_{\pm }\left( \beta ,\theta \right) &=&\exp \left\{ -\beta \hbar \tilde{%
\omega}\left( 1\pm \frac{m\omega }{2}\theta \right) \right\}
\sum_{n^{+}}\exp \left\{ -\beta \hbar \tilde{\omega}\left( 1+\frac{m\omega }{%
2}\theta \right) n^{+}\right\} \times  \notag \\
&&\sum_{n_{+}}\exp \left\{ -\beta \hbar \tilde{\omega}\left( 1-\frac{m\omega 
}{2}\theta \right) n_{+}\right\}  \notag \\
&=&\frac{\exp \left\{ \mp \beta \hbar \frac{m\omega ^{2}}{2}\theta \right\} 
}{4\sinh \beta \hbar \tilde{\omega}\left( 1+\frac{m\omega }{2}\theta \right)
\sinh \beta \hbar \tilde{\omega}\left( 1-\frac{m\omega }{2}\theta \right) } 
\notag \\
&=&\exp \left\{ \mp \beta \hbar \frac{m\omega ^{2}}{2}\theta \right\}
Z_{HO}\left( \beta ,\theta \right),
\end{eqnarray}
where 
\begin{equation}
Z_{HO}\left( \beta ,\theta \right) =\frac{1}{4\sinh \beta \hbar \tilde{\omega%
}\left( 1+\frac{m\omega }{2}\theta \right) \sinh \beta \hbar \tilde{\omega}%
\left( 1-\frac{m\omega }{2}\theta \right) }
\end{equation}
is the partition function of the harmonic oscillator in non-commutative
space $\left[ 14,15 \right] $.

Now we are in position to compute several interesting quantities. The Free
energy of a systems at finite temperature is: 
\begin{eqnarray}
F_{\pm }\left( \beta ,\theta \right) &=&-\frac{1}{\beta }\ln Z_{\pm }\left(
\beta ,\theta \right)  \notag \\
&=&\pm \frac{\hbar m\omega ^{2}}{2}\theta -\frac{1}{\beta }\ln Z_{HO}\left(
\beta ,\theta \right),
\end{eqnarray}
which at low temperature limit tends to the ground state energy of system.
In this case we have: 
\begin{equation}
\lim_{\beta \rightarrow \infty }F\left( \beta ,\theta \right) =\hbar \tilde{%
\omega}\left( 1\pm \frac{m\omega }{2}\theta \right),
\end{equation}
which corresponds to ground state energy of the system, $E_{00}=\hbar \tilde{
\omega}\left( 1\pm \frac{m\omega }{2}\theta \right)$. The mean energy of a
systems is: 
\begin{eqnarray}
\left\langle E_{\pm }\left( \theta ,\beta \right) \right\rangle &=&-\frac{%
\partial }{\partial \beta }\ln Z_{\pm }\left( \beta ,\theta \right)  \notag
\\
&=&\mp \frac{\hbar m\omega ^{2}}{2}\theta +\hbar \tilde{\omega}\left[ \left(
1+\frac{m\omega }{2}\theta \right) \cosh \beta \hbar \tilde{\omega}\left( 1+%
\frac{m\omega }{2}\theta \right) \right.  \notag \\
&&\left. +\left( 1-\frac{m\omega }{2}\theta \right) \cosh \beta \hbar \tilde{%
\omega}\left( 1-\frac{m\omega }{2}\theta \right) \right].
\end{eqnarray}

Let us now investigate the infinite temperature limit $\left( \beta
\rightarrow 0\right) $ where $\theta $ and $\omega $ are fixed. At this
limit Eq. (51) collapses into: 
\begin{equation}
\lim_{\beta \rightarrow 0}\left\langle E_{\pm }\left( \theta ,\beta \right)
\right\rangle =\pm \frac{\hbar m\omega ^{2}}{2}\theta +2\hbar \tilde{\omega}%
\frac{kT}{\hbar \tilde{\omega}}=\hbar \tilde{\omega}\left( \frac{2kT }{\hbar 
\tilde{\omega}}\pm \frac{m\omega }{2}\theta \right).
\end{equation}
This means that the non-commutativity effects are manifested in energy at
the high temperature limit, contrary to what was found in the reference $%
\left[ 15\right]$.

\section{ Klein-Gordon Oscillator in Non-commutative complex space}

The Klein-Gordon equation in complex quantum space and in constant magnitic
field has the following form: 
\begin{equation}
\left( 2p_{\hat{z}}-ie\frac{B}{2}z\right) \left( 2p_{z}+ie\frac{B}{2}\bar{z}
\right) \psi =\left( E^{2}-m^{2}\right) \psi.
\end{equation}
which can be written in commutative space as: 
\begin{equation}
\left( p_{x}^{2}+p_{y}^{2}+\frac{e^{2}B^{2}}{4}\left( x^{2}+y^{2}\right)
-eBL_{z}\right) \psi =\left( E^{2}-m^{2}+eB\right) \psi.
\end{equation}
This equation is same as the Klein-Gordon equation in real space and in
magnetic field with the extra constant ($eB$). In a non-commutative complex
quantum space, it is described by the following equation: 
\begin{multline}
\left( 
\begin{array}{cc}
\left( 2p_{z}+ie\frac{B}{2}\hat{\bar{z}}\right) \left( 2p_{\hat{z}}-ie \frac{%
B}{2}\hat{z}\right) & 0 \\ 
0 & \left( 2p_{\hat{z}}-ie\frac{B}{2}\hat{z}\right) \left( 2p_{z}+ie\frac{B}{
2}\hat{\bar{z}}\right)%
\end{array}
\right) \psi = \\
= \left( E^{2}-m^{2}\right) \psi.
\end{multline}

Using the definition of the non-commutative coordinates, we can rewrite this
equation in a commutative quantum space as: 
\begin{multline}
\left( (1+\frac{eB\theta }{4})^{2}\left( p_{x}^{2}+p_{y}^{2}\right) +\frac{%
e^{2}B^{2}}{4}\left( x^{2}+y^{2}\right) \right. \\
\left.-eBL_{z}-\frac{eB\theta }{4}\left( L_{z}\pm 2\left( \frac{1}{2}\right)
\right) \right) \psi = \left( E^{2}-m^{2}+eB\right) \psi,
\end{multline}
so that the critical point is obtained when the coefficient of the extra
non-commutative constant equals to zero. In this case the non-commutativity
parameter $\theta =\frac{4}{eB }$. This value is the same one obtained in
the case of the equivalence between non-commutative quantum mechanics and
the Landau problem $\left[ 14\right]$. At this point it is clear that the
equation $(56)$ is similar to the Klein-Gordon one in a commutative space
with an external magnetic field.

The Klein-Gordon oscillator in complex space can be defiend by the following
equation: 
\begin{equation}
\left( 2p_{\bar{z}}+im\omega z\right) \left( 2p_{z}-im\omega \bar{z}\right)
\psi =\left( E^{2}-m^{2}\right) \psi,
\end{equation}
which can be rewritten in commutative space as: 
\begin{equation}
\left( p_{x}^{2}+p_{y}^{2}+m^{2}\omega ^{2}\left( x^{2}+y^{2}\right)
+2m\omega L_{z}\right) \psi =\left( E^{2}-m^{2}+2m\omega \right) \psi.
\end{equation}
It is clear that the K-G oscillator in a complex space is similar to the K-G
equation for a particle of positive charge in an external magnetic field.
The eigenvalues for the Hamiltonian in equation $(58)$ are: 
\begin{eqnarray}
E^{2} &=&2m\omega \left( n_{x}+n_{y}+1\right) +2m\omega \left(
m_{l}-1\right) +m^{2}  \notag \\
&=&2m\omega \left( n_{x}+n_{y}+m_{l}\right) +m^{2}.
\end{eqnarray}
In the non-commutative complex space the K-G oscillator is described by the
following equation: 
\begin{multline}
\left( 
\begin{array}{cc}
\left( 2p_{z}+im\omega \hat{\bar{z}}\right) \left( 2p_{\bar{z}}-im\omega 
\hat{z}\right) & 0 \\ 
0 & \left( 2p_{\bar{z}}-im\omega \hat{z}\right) \left( 2p_{z}+im\omega \hat{%
\bar{z}}\right)%
\end{array}
\right) \psi = \\
=\left( E^{2}-m^{2}\right) \psi.
\end{multline}
Using the relations $(6$) and $(7)$ we can rewrite the equation $(60)$ in
commutative space as: 
\begin{multline}
\left( (1-m\omega \theta +\frac{m^{2}\omega ^{2}}{4}\theta ^{2})\left(
p_{x}^{2}+p_{y}^{2}\right) +m^{2}\omega ^{2}\left( x^{2}+y^{2}\right)
+\right. \\
+ 2m\omega L_{z}-m^{2}\omega ^{2}\theta \left( L_{z}\pm 1\right) \bigg) \psi
=\left( E^{2}-m^{2}+2m\omega \right) \psi.
\end{multline}
This equation is similar to the equation of motion for a fermion of spin $%
\frac{1}{2}$ in a constant magnetic field. Then the equation $(61)$ takes
the following form: 
\begin{multline}
\left( (1+\frac{m\omega \theta }{2})^{2}\left( p_{x}^{2}+p_{y}^{2}\right)
+m^{2}\omega ^{2}\left( x^{2}+y^{2}\right) -2m\omega L_{z}\right. \\
-m^{2}\omega ^{2}\theta \left( L_{z}+2s_{z}\right) \bigg) \psi =\left(
E^{2}-m^{2}-2m\omega \right) \psi ,\quad s_{z}=\pm \frac{1}{2},
\end{multline}
where the energy eigenvalues for eq. $(62)$ are given by: 
\begin{equation}
E^{2}=2m\omega _{\theta }\left( n_{x}+n_{y}+1\right) +2m\omega _{\theta
}\left( m_{l}\pm 1\right) +m^{2},
\end{equation}
where $\omega _{\theta }=\omega (1-\frac{m\omega \theta }{2})$. We have
found that the non-commutativity plays the role of a magnetic field
interacting automatically with the spin of a particle, thereby the system
with spin in a magnetic field will have a resonance $\left[ 17\right]$. Then
the critical values of $\theta =\frac{2}{m\omega }$ can be considered as a
resonance point.

\section{Conclusion}

In this work we started from quantum charged oscillator in a canonical
non-commutative complex space. By using the Moyal product up to first order
in the non-commutativity parameter $\theta $, we derived the deformed
Hamiltonian and Klein-Gordon equation. By solving it exactly we found that
the energy is shifted up to the first order in $\theta $ by two levels,
hence we can say that the particle in non-commutative complex space is
equivalent to a particle with spin $1/2$ in magnetic field, where the
non-commutativity plays the role of magnetic field which creates the total
magnetic moment of particle with spin $1/2$, which in turn shifts the
spectrum of energy. Such effects are similar to the Zeeman splitting in a
commutative space.

\end{document}